\documentstyle[]{laa}

\newcommand{\be}{\begin{equation}}
\newcommand{\ee}{\end{equation}}
\newcommand{\bea}{\begin{eqnarray}}
\newcommand{\eea}{\end{eqnarray}}

\begin{document}
\title{Polarisation as a Tool for Gravitational Microlensing Surveys}
\author{John F.L. Simmons\inst{1}
  \and Jon P. Willis\inst{1}
  \and Andrew M. Newsam\inst{1}}
\offprints{J. Simmons}
\institute{Department of Physics and Astronomy, University of Glasgow}
\date{Received date; accepted date}
\maketitle
\begin{abstract}
Much interest has been generated recently by
the ongoing MACHO, EROS and OGLE projects to identify
gravitationally lensed stars from the Large Magellanic Cloud
and Galactic bulge, and the positive identification of
several events 
(Alcock et al, (\cite{Alcock+93}), Aubourg et al, (\cite{Aubourg+93}) and
Udalski et al, (\cite{Udalski+93})).
The rate at which such
events are found should provide considerable information
about the distribution of the low mass compact objects,
brown dwarfs etc. responsible for the lensing, and
their relative importance as a component
of the cosmological `dark matter'. We show
measurement of the polarisation of starlight during these
events can yield considerably more information about
the lensing objects than was previously considered
possible. Furthermore, the consideration of extended sources
is shown to have a significant effect on the interpretation
of the profiles and statistics of the events.
\end{abstract}

More detailed modelling of the gravitational microlensing of
stars by low mass objects can yield considerable information about
the lens and the lensing geometry. In particular, the
measurement of the variable polarisation produced by
gravitational lensing of stars could in principle provide
further information about the mass, distance and velocity of the
lensing objects.
This possibility appears to have been overlooked in both the
MACHOs and EROS programmes.

Although the idea that polarisation could be produced by
gravitational lensing was raised and investigated in
relation to supernovae (Schneider and Wagoner, \cite{Schneider+87}), 
it seems that not to have been
considered for stars. Polarisation
produced by lensing of stars is not large, but should be
measurable for sufficiently bright stars. 

\begin{figure}[thc]
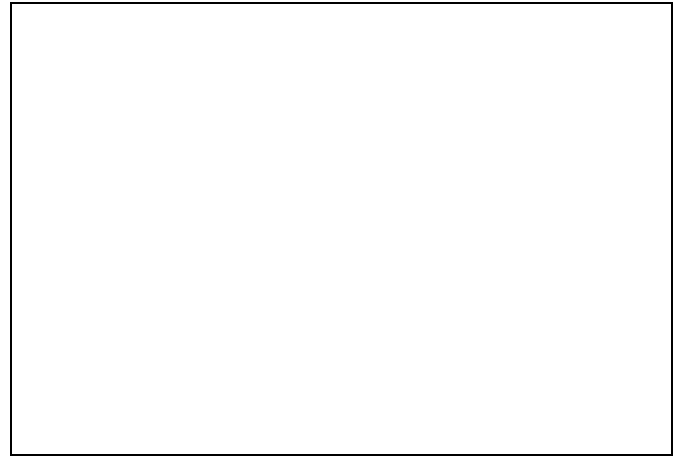

  \picplace{6cm}
\caption[]{
Two stages during an event. In (a), the event is about to begin. There is
no noticeable difference in the amplification of the N/S and E/W regions so the
star remains unpolarised (although a flux increase might already have been 
seen). In (b), however, the lens is at closest approach and the significant
amplification of South w.r.t. East and West gives a net polarisation in 
the direction at S. 
\label{fig:event}}
\end{figure}
Normally the light received from stars is unpolarised,
unless the star is rotationally distorted, or has an
asymmetric envelope or wind. However the light that emerges
from the limb of a star can be expected to be polarised (up
to 10\%). This effect, predicted by Chandrasekhar (\cite{Chand60}) who
calculated it for an electron scattering photosphere,
depends on the scattering mechanism and on the direction of
the emergent radiation from the normal to the stellar
photosphere. It has been observed in the Sun. However, for a
star, the polarised light flux observed at the Earth is
obtained by integrating the polarised intensity over the
stellar disc. The polarisation of the light at the
north/south points is in an opposite direction to that at
the east/west points (see figure~\ref{fig:event}).
If the star is rotationally symmetric, no net polarisation
should be observed owing to cancellation. In the case of a
gravitationally lensed star, the amplification at different
points on the stellar disc will be different, so the net
polarisation differs from zero by an amount depending on the
distance of the lensing object projected onto the source
(star's) plane. The direction of this polarisation will also
vary with as angle of the line of centres of source and
lensing object changes.

One can obtain an estimate of the degree of
polarisation for a Schwarzschild lens as follows. 
Distances are most conveniently written in units of
the Einstein radius projected onto the source plane, $\eta_0$, 
given by
\be
{\eta_0}^2=\left ({1 \over {a_L} }-{1 \over {a_S}}\right ) 2 R_S {a_S}^2
\ee
$a_L$ and $a_S$ are the distances to the lens and source respectively, 
and $R_S={{2GM} /{c^2}}$ is the Schwarzschild radius of the lensing object. 
The Amplification at distance $x$ is given by
$A(z)={1 \over 2} \left ( z + {1 \over z}\right )$ 
where $z=\left ( 1+{4 \over x^2} \right )^{1/2}$.
For small $x$, $A(x) \sim {1/x}$. For polarisation greater
that 1\%, say, this amplification must vary by 10\% over one
stellar radius, $R$ (assuming a limb polarisation of around
10\%). This immediately yields the condition that $R>0.1 d$,
where $d$ is the distance of approach to the centre of the
stellar disc.

Exact expressions for the polarised and unpolarised flux,
$F_U$ ,$F_Q$ and $F_I$ can readily be obtained from the form
of the polarised and unpolarised intensity (stokes
parameters) as functions of the angle of emergence to the
normal, $\theta$, at the surface of the star. If we assume these to
take the form
\bea
I&=&i_0 +i_1 \mu\label{eq:Inten}\\
U&=&u_0(1-\mu )
\eea
where $\mu = \cos\theta$,
we obtain on integrating over the stellar disc the
expressions
\bea
F_I &=& {{R^2} \over {{a_s}^2}} \int_0^{2\pi} \int_0^1 (i_0 +i_1
\mu )\ A(x(\mu,\phi ))\mu\, d\mu d\phi\\
F_U &=& {{R^2} \over {{a_s}^2}} \int_0^{2\pi} \int_0^1 u_0(1-\mu
)\ A(x(\mu,\phi ))\mu \,  \cos {2\phi }\ d\mu d\phi\\
F_Q &=& {{R^2} \over {{a_s}^2}} \int_0^{2\pi} \int_0^1 u_0(1-\mu
)\ A(x(\mu,\phi )) \mu \, \sin {2\phi }\ d\mu d\phi
\eea
where 
\be
x^2=d^2 + R^2 (1-\mu^2) -2 (1-\mu^2)^{1/2}R d
\cos \phi
\ee
$a_s$ is the distance to the source and 
$\phi $ the position angle.

The observed degree of polarisation is given by
\be
p={{ ({F_U^2 +F_Q^2})^{1 \over 2}} \over F_I}
\ee
and is a
function of $R$ and $d$. (In fact $F_Q=0$ in the coordinate
system chosen). The latter is easily expressed in terms of
the transit velocity of the lensing object and the distance
of closest approach or impact parameter, $d_0$ and time (all
in the source plane).

Thus simultaneous measurement of the time variable
polarisation and unpolarised flux yields a lot of
information about the lensing set-up. If one assumes the
radius of the star is known, then one immediately obtains
the value of $\eta_0$. Modelling of the stellar atmosphere
should not even be necessary, as the parameters $i_0$, $i_1$
and $u_0$ could be obtained from the time profile of $p$,
$F_I$ and position angle. If one assumes additionally, 
though less reasonably,a
transit velocity for the lensing object one can obtain the
distance to the lens, and indeed its Schwarzchild radius
(mass). (Of course one does need to assume a distance to the
source).

The statistics of the events are potentially more important.
We may consider three types of event
{\setlength{\parsep}{0pt}
\itemsep=0pt
\begin{enumerate}
\item flux variation
\item polarisation variation
\item  transit events (i.e. where some part of the source is
directly in line with the lens centre and observer) which
should appear as sudden increases in flux.
\end{enumerate}}

An event (i) will be recorded when the flux variation is
greater than some chosen value. In the case of a point
source, this will simply be when the impact parameter is
less than some constant times the Einstein radius, $\eta_0$,
i.e. $d_0 < \alpha \eta_0$. The behaviour of the time
profile of the flux for an extended source is significantly
different to that of a point source (Simmons et al, \cite{Simmons+94}). 
If $R \sim d_0$
then the amplification is smaller in the wings but higher at
closest approach, giving a sharper profile. This is possibly
the explanation of the outlier at maximum amplification
observed by the MACHOs group for the first event reported
(Alcock et al, \cite{Alcock+93}). 
On the other hand if $R>3$ the amplification is
almost unobservable regardless of the impact parameter.
Since the Einstein radius, $\eta_0$ depends on $M^{1/2}$,
this would seriously impede the detection of low mass
lenses.

Cases (ii) and (iii) will only occur for extended sources.
The polarisation time profile in (iii) will be quite
specific, with two equal maxima, and a central minimum. The
relative frequency of the different types of events will
indicate the spatial distribution of lensing objects.

\begin{figure}[thc]
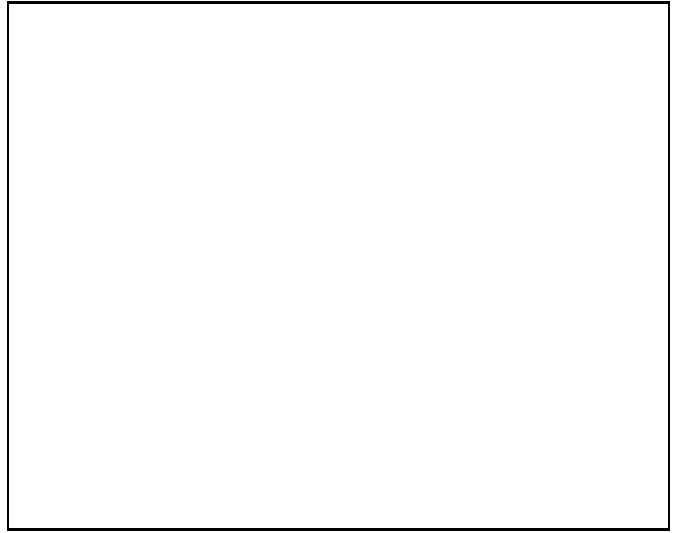

  \picplace{7cm}
\caption[]{
Contours of 0.1\% polarisation and a flux amplification of 1.34 
(corresponding to amplification of a point source at the Einstein radius) as
a function of source radius and impact parameter. 
For values of $R$ and $d_0$ inside the polarisation
``ellipse'' variable polarisation will be seen. Similarly, points
in the bottom left corner enclosed by the flux contour are flux events. 
The points with $R>d_0$ are transit events.
\label{fig:Domain}}
\end{figure}
The domain of values in the $R$ and $d$ plane is given in
Figure ~\ref{fig:Domain} (see also Simmons et al, \cite{Simmons+94}), 
the unit of distance being $\eta_0$. Thus
for lenses of the same mass near the galactic centre $R$ and
$d$ take small values, and for lenses near the LMC $R$ and
$d$ are large. If the stars in the source plane are taken to
be uniformly and randomly distributed, the fraction of
events of any one type for a fixed radius of star is simply
proportional to the length of the interval in $d$. For example, for
$R=1$, more than half of flux events also show variable polarisation
(see figure). 

From the rate of events, and the relative frequency of types
(i), (ii) and (iii) it should be possible, given a
sufficiently large sample, determine the spatial
distribution of lenses.

Two further points should be stressed. The rise in
polarisation takes place later than the rise in total flux,
by approximately a factor of 2. Thus an event suspected
because of an observed amplification in flux, could be
monitored and confirmed polarimetrically. Secondly, the
polarisation effects arise from the limb dependency of
polarisation in an extended source. Even for an unpolarised
source the parameters occurring in expression (\ref{eq:Inten}) for the
intensity should depend on wavelength both in continuum and
lines, so one might expect that for an extended source where
limb darkening is present the relative amplitude of the flux
variation would depend on frequency. For a point source, or
for a star with no limb darkening, this effect would not be
seen, and the profile would be achromatic.

Polarisation measurements are certainly feasible.
Interstellar polarisation might make interpretation more
difficult, but as we are dealing here with variable
polarisation, this could be overcome. This should also
provide valuable information about the lensing, and the
importance of MACHOs as a dark matter component.

\section{Acknowledgements}

JPW performed this work as part of a Cormack Research Scholarship. AMN
was funded by a PPARC Postgraduate Studentship.

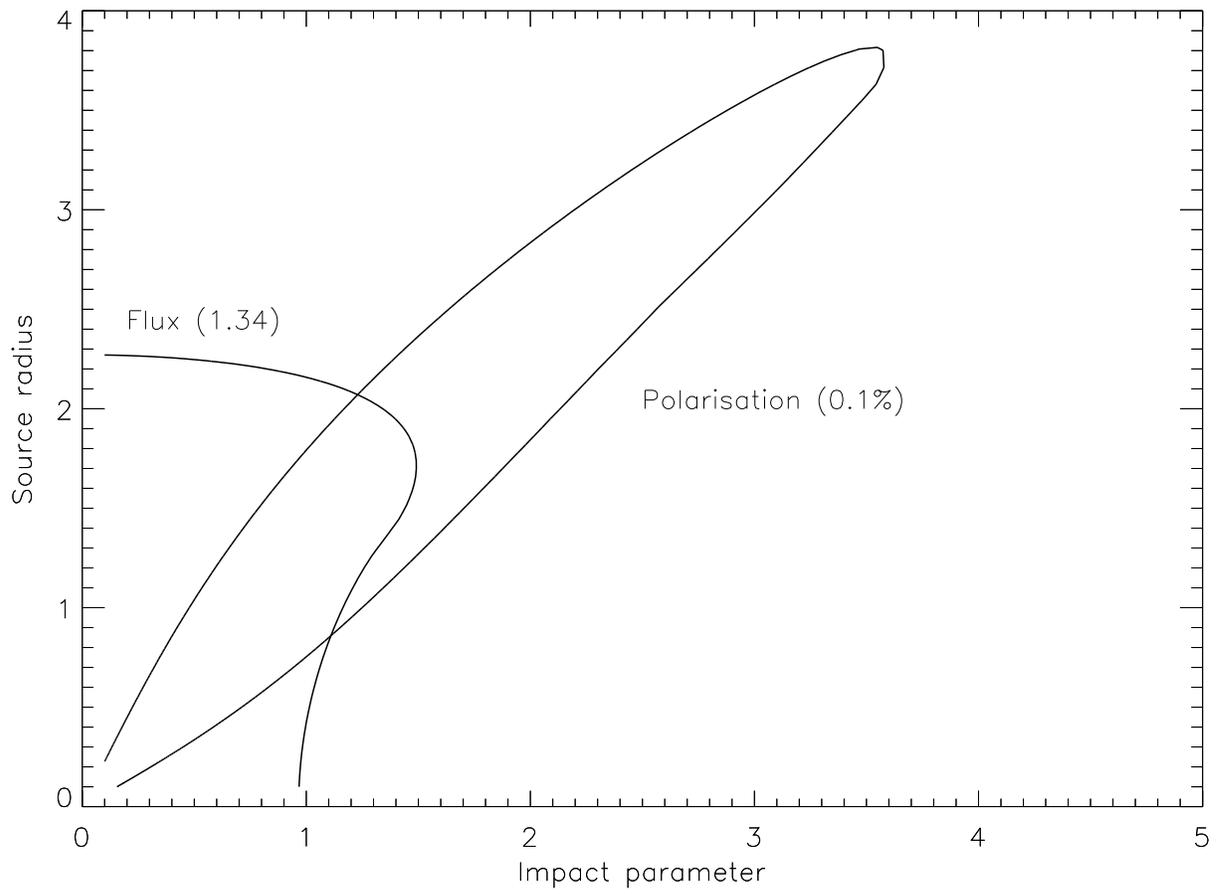
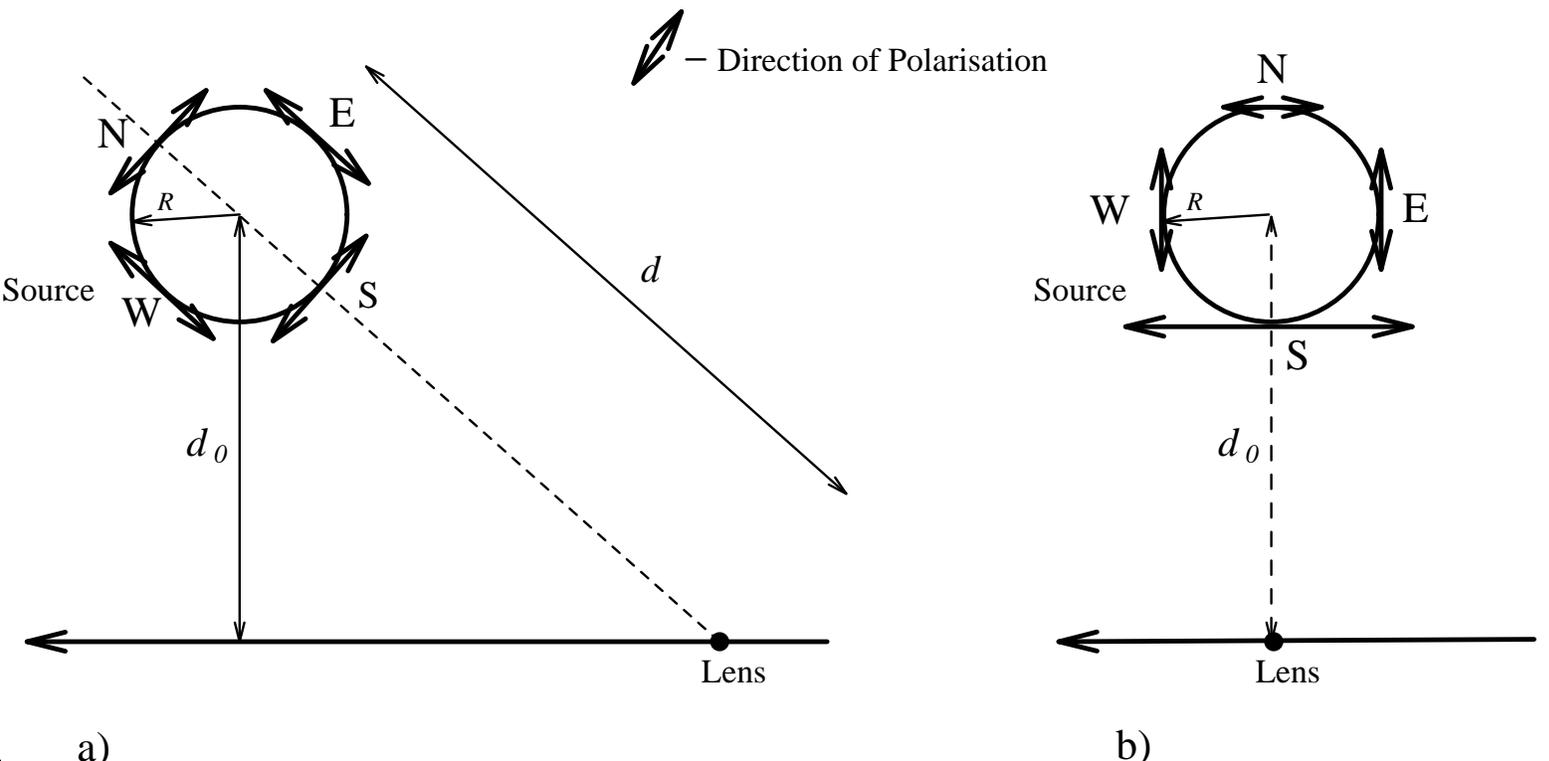
\end{document}